# Plasmonic lasing of nanocavity embedding in metallic nanoantenna array

Cheng Zhang [†], Yonghua Lu[†※], Yuan Ni[†], Mingzhuo Li[†], Lei Mao[†], Chen Liu[†],

Douguo Zhang[†], Hai Ming[†] and Pei Wang[†※]

[†]Department of Optics and Optical Engineering, Anhui Key Laboratory of

Optoelectronic Science and Technology, University of Science and Technology of

China, Hefei, Anhui 230026, China.

**ABSTRACT:** Plasmonic nanolasers have ultrahigh lasing thresholds, especially those devices for which all three dimensions are truly subwavelength. Because of a momentum mismatch between the propagating light and localized optical field of the subwavelength nanocavity, poor optical pumping efficiency is another important reason for the ultrahigh threshold but is normally always ignored. Based on a cavity-embedded nanoantenna array design, we demonstrate a room-temperature low-threshold plasmonic nanolaser that is robust, reproducible, and easy-to-fabricate using chemical-template lithography. The mode volume of the device is~$0.22(\lambda/2n)^3$ (here, $\lambda$ is resonant wavelength and n is the refractive index), and the experimental lasing threshold produced is ~2.70MW/mm$^2$. The lasing polarization and the function of nanoantenna array are investigated in detail. Our work provides a new strategy to achieve room-temperature low-threshold plasmonic nanolasers of interest in applications to biological sensing and information technology.

Subwavelength confinement of electromagnetic fields in plasmonic lasers enhances significantly light-matter interactions, and opens up a wide range of applications in spectroscopy, bio-sensing, imaging and information technology[1-5]. Proposed ten years ago and referred to as a 'spaser'[6, 7], the plasmonic nanolaser generates stimulated emission of surface plasmons in the same way as a laser radiating coherent light. Thereafter, spasers have been experimentally demonstrated in different metallic nanostructures, such as, gold colloids surrounded by dye-shells[8, 9], hybrid plasmonic waveguides based on nanowires[10-15], metallic-coated nanocavities[16], metal nanopan cavities[17] and plasmonic nanocavity arrays[18-20]. Ultrahigh plasmonic loss is still the main obstacle to its development and application[21]. Most research struggle to overcome plasmonic loss by new cavity designs or improvements in fabrication. The strategy possibly works for a plasmonic laser[10-13] that constrains only two dimensions of the device to subwavelength levels. For a truly three-dimensional (3D) subwavelength-constraint spaser[8, 9] with localized surface plasmons, the lasing threshold is still very high, even for plasmonic nanocavity arrays[18-20].

Regarding conventional lasers, the experimental lasing threshold is known to be determined by not only the intrinsic loss of the cavity but also the pumping efficiency, which has drawn little attention in previous reports on spasers. Although electric pumping is considered to offer better prospects for spasers[22, 23], optically pumping is believed to be irreplaceable in some specific applications such as high-speed all-optical signal processing and ultrafast spectroscopy[24]. Therefore, most experimental reports have to date focused on optical pumping. The gap between the wavelength of the pump beam and the subwavelength size of the spaser device hinder the energy exchange between them; this is a well-known fundamental problem in nano-optics[25]. Optical antennas have been proposed and proved to be efficient devices to convert free-propagating optical radiation into localized energy[26]. Determined by their geometry, plasmonic nanoantennas can be designed to resonate at the pumping wavelength, emission wavelength or both[27, 28]. If the antennas are arrayed, subradiant plasmons induce a stronger absorption and field enhancement[29], and the antenna array can also tailor the radiation into a well-defined field pattern[18]. Hence, lowering the experimental lasing threshold of a spaser seems promising because combining plasmonic nanoantenna with nanocavity will refuel the pumping energy into the nanocavity more efficiently.

Here, we report a low-threshold, room-temperature plasmonic laser and demonstrate that an optical antenna array can efficiently lower the lasing threshold of a spaser. The new device is constructed by embedding a fluorescence polystyrene sphere into a silver nanoparticle (nanoantenna) array, which combines the nanocavity and optical antennas together to promote pumping efficiency. The lasing threshold is 2.70 MW/mm$^2$, more than 20 times lower than that of a room-temperature arrayed nanocavity spaser[18]. Unlike earlier struggles to decrease lasing thresholds by eliminating spaser intrinsic metal loss, we propose to increase the pumping efficiency with the optical antenna array. This array will resonantly absorb light from the pump beam and concentrate the energy into the cavity of the embedded fluorescence polystyrene sphere. Moreover, the cavity mode, which is mainly localized within the polystyrene sphere, will also reduce the intrinsic metal loss. The lasing threshold and other properties of the spaser are reported and their dependences on geometric parameters of the devices are described.

The lasing system, sketched in Fig.1a, is constructed by embedding a fluorescence polystyrene ball into a two-dimensional silver nanoantenna array. The 150-nm-diameter ball functions not only as a gain medium but also as a nanocavity together with the surrounding silver nanoantennas. A scanning electron microscopy (SEM) micrograph (Fig.1b), shows a periodicity of about 65 nm for the silver nanoantenna array and a diameter of about 100 nm for the cavity. The height of the nanoantenna is around 15 nm as indicated in the AFM image (Fig.1c). All these parameters are optimized to match the antenna and cavity modes with the pumping and lasing frequencies, respectively. The far-field reflectivity spectrum (Fig.1f) measured around the nanocavity indicates a shallow absorption resonance dip at around 530 nm corresponding to the mode of the nanoantenna array and a sharper dip around 620 nm for the cavity mode (the modes were assigned by comparing the spectrum with that of a sample without the nanocavity (Supplementary Fig.S1). Both the antenna mode and the cavity mode are simulated using 3D finite difference time domain (FDTD) modelling under optimized geometrical parameters. The calculated reflectivity spectrum (Supplementary Fig.S2) exhibits two resonance dips that match our measurements very well. The antenna mode presented in Fig.1d indicates that the pump-beam light (far field) is concentrated efficiently into the subwavelength cavity with the aid of the nanoantenna array. The cavity mode (Fig.1e) for lasing is well confined within the polystyrene ball through localized surface-plasmon resonance of the embedded cavity. The mode volume is estimated to be $0.22(\lambda/2n)^3$ for the 647-nm cavity mode (Supplementary Fig.S3) corresponding to a high local density of optical states.

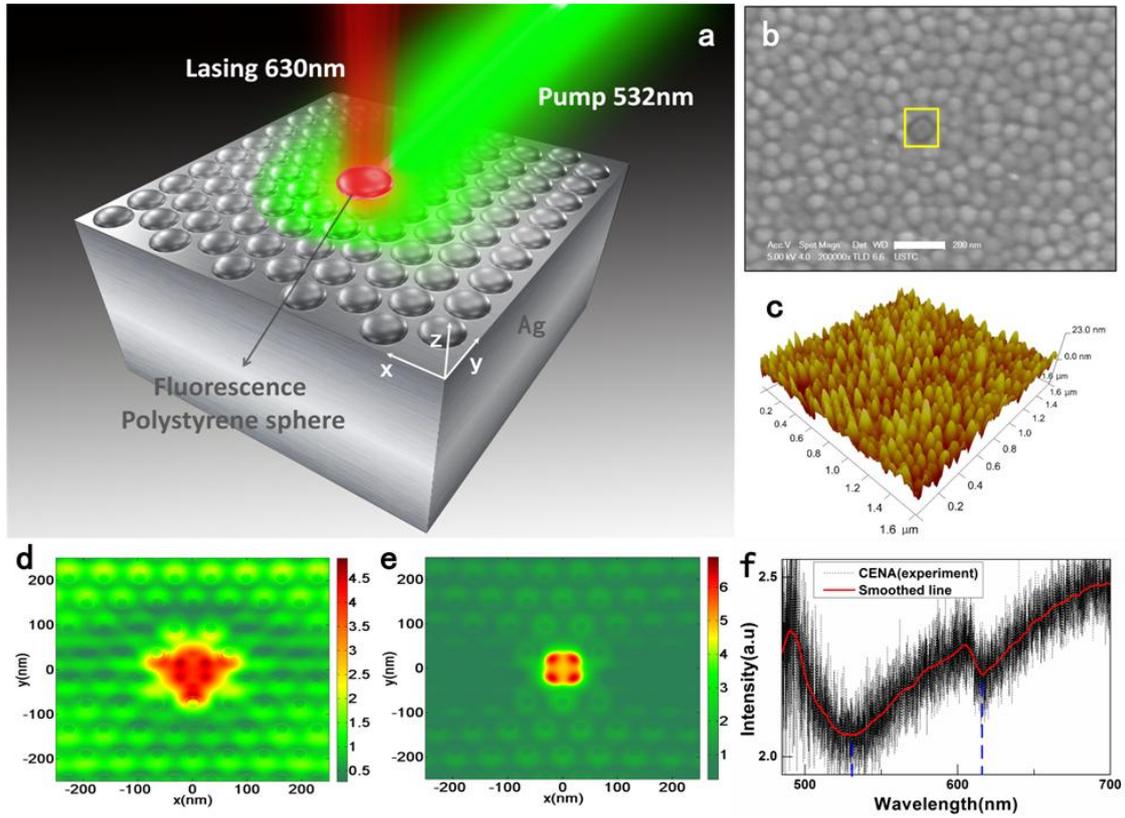

**Figure 1. Design of the cavity-embedded nanoantenna array (CENA) device. (a)** Schematic of the lasing system for the CENA device, **(b)** Top view SEM image for the CENA device with its scale bar of 200 nm (the embedded nanocavity is marked by the yellow square). **(c)** 3D view by AFM for an antenna array (without embedded nanocavity) with periodicity of 65 nm. The scan range here is 1.6 μm×1.6 μm. **(d-e)** 3D FDTD numerical simulations of the electric field distributions for wavelengths 532 nm (antenna mode) and 645 nm (cavity mode). **(f)** Far field reflection spectra measurement result for the CENA device. The experimental data are presented as the black dotted line along with its smoothed line in red.

Combining the antenna and cavity modes, the CENA device radiates coherently under optical pumping using a 532-nm pulsed laser beam at room temperature (details of the measurement is presented in Supplementary). Fig.2a presents the lasing spectra at different pumping energies; the spectrum is just the photoluminescence at a low pumping energy (~1 µJ). Lasing emerges after reaching the pumping threshold of 2.70 MW/mm$^2$, which is only 4%–5% that of other room temperature plasmonic nanolaser devices[18]; the lasing emission peak centred at 630 nm shrinks to a narrow peak with full width at half maximum (FWHM) of 4.3 nm. Figure 2b gives the output lasing power (integrating stimulated emission intensity from 627 nm to 633 nm) and FWHM as a function of pumping intensity in a double-logarithmic plot. The S-shaped curve representing the transition from photoluminescence to lasing regime matches the fitted line based on the threshold theory for single-mode laser oscillations[30]. From the threshold theory, the spontaneous emission input into the laser can be quantified using parameter $x_0'$ [30], which takes very small values for conventional lasers, for example ~$10^{-8}$ for Nd:YAG[30], but is much larger for a microcavity laser because of its ultrahigh mode confinement[31]. For our CENA device $x_0'$ is 0.0136, which indicates the lasing mode is well confined.

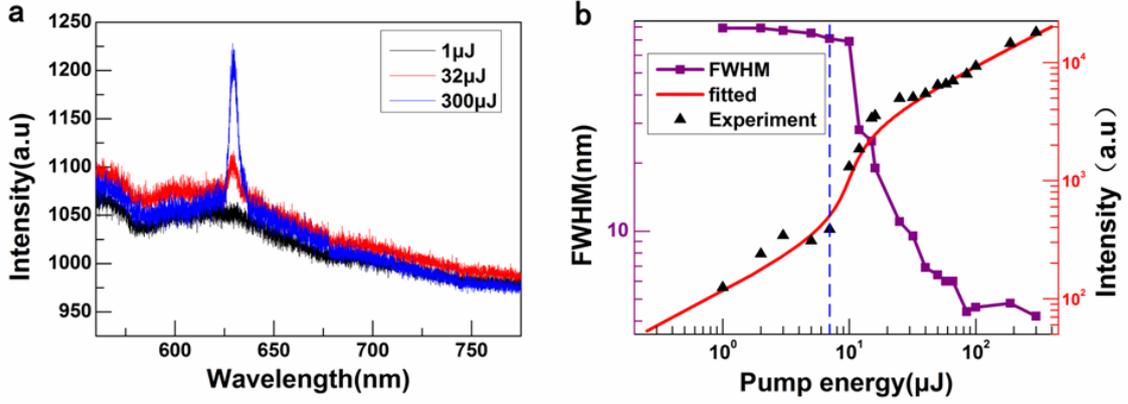

**Figure 2. Lasing results and threshold features for the CENA device. (a)** The three spectra for different pump intensities from spontaneous emission one (black line, pump intensity ~1 μJ) to a full laser oscillation (navy blue line, pump intensity ~300 μJ). **(b)** Output power responses as a function of pump energy with the experimental data (black dots) well fitted (red line) using parameter value $x_0^{'} = 0.0136$. The parameter of $x_0^{'}$ is the ratio of the spontaneous emission input into laser. With increasing input energy, FWHM rapidly decreases from 80 nm to 4.3 nm.

Plasmonic lasers have specific polarization characteristics arising from surface plasmon[32]. We compared the spectra of polarized radiation from our CENA device with that of an isolated cavity embedded in a silver film by inserting a linear polarizer in front of the spectrometer (Supplementary Fig.S8). By rotating the polarizer by 90°, polarization-dependent spectra were obtained for the two orthogonal polarizations for the CENA device (Fig.3a-b) and for the cavity only device (Fig. 3c-d). The lasing peak in the spectra is unambiguously polarized and the spontaneous radiation is unpolarized. By integrating over the lasing frequencies from 627 nm to 633 nm, the polarization pattern (red dotted line in Fig.3e) exhibits clear characteristic of linear-polarization resulting from surface plasmons mediating the lasing process[17]. The pattern for the cavity-only device (black dotted line in Fig.3e) corresponds to non-polarized emissions indicating that only spontaneous radiation is recorded even for the highest optical pumping energies in our experiments.

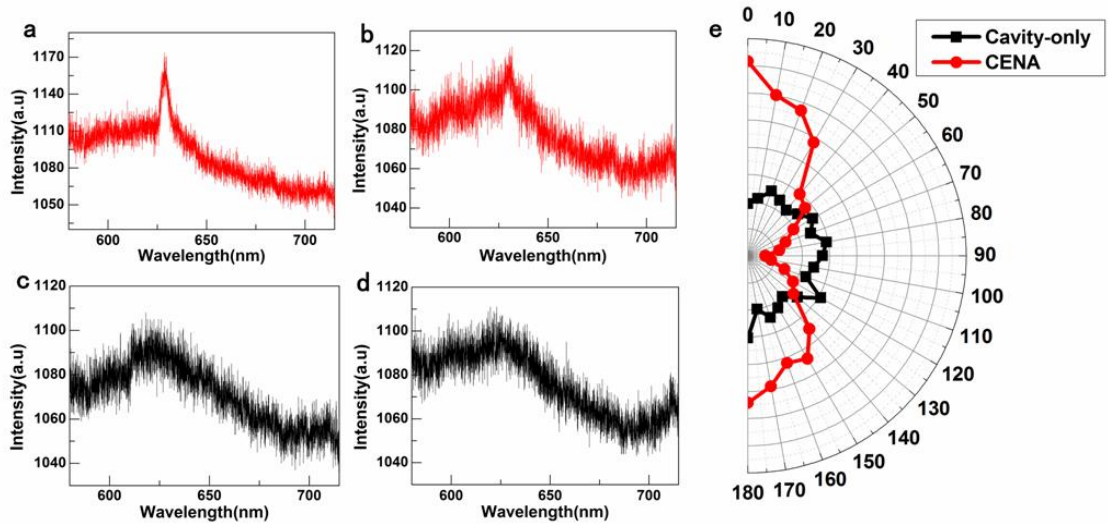

**Figure 3. Characteristics of lasing polarization state for CENA and cavity-only devices. (a, b)** CENA emission spectra with polarizer angle of 0° and 90°. **(c, d)** emission spectra of cavity-only with polarizer angle of 0° and 90°. **(e)** Experimental output power of the lasing mode as a function of polarization angle.

The nanoantenna, the key element in a CENA device, elevates the optical pumping efficiency. To understand the function of the nanoantenna array in the CENA device, another CENA sample (CENA-2) with a different periodicity (155 nm) and height (90 nm) (Supplementary Fig. S4) was studied for comparison in Fig.4a. Shifted to 585 nm (Fig. 4b), the antenna mode did not resonate with the pumping laser (532 nm), hence the pumping efficiency decreased dramatically. As shown in Fig. 4d (red dotted line), the output power as a function of pump energy indicates a threshold at 97.5 μJ (~37.64 MW/mm$^2$), which is 14 times higher than that of the above optimal CENA sample. The centre wavelength for lasing is still located at 630 nm (Fig.4a) with a slight broadening in FWHM (7 nm) as the size of the cavity is approximately the same as that of the CENA device. Furthermore, comparing with the cavity-only sample in which only a single cavity mode at 645 nm exists (Supplemental Fig. S5), no antenna mode can be excited in this sample; this clarifies the function of the antenna mode. The emission spectra at different pumping energies and the threshold curves (the output power as the function of pump energy) in Fig. 4c-d suggest that no lasing occurs for pump energies up to 330 μJ, the highest energy used in our experiments.

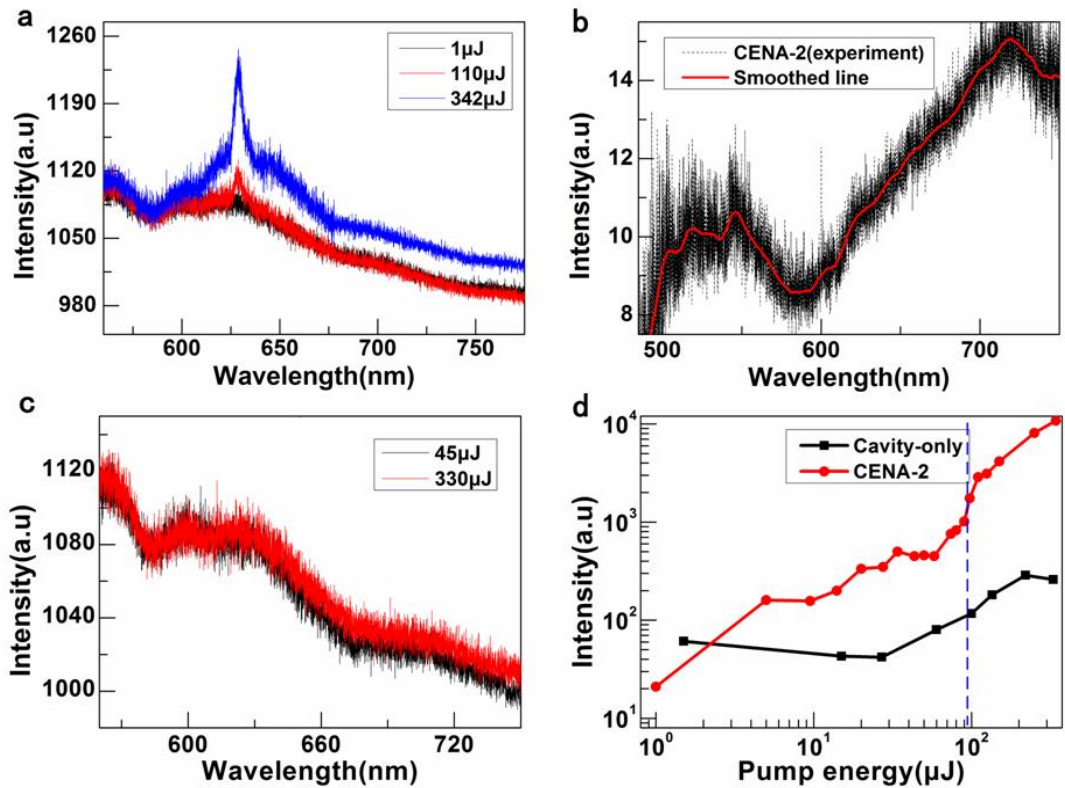

**Figure 4. Comparison of lasing results.** (a) Three emission spectra with input

energy of 1 μJ, 110 μJ and 342 μJ. **(b)** Far-field reflection spectrum of CENA-2. **(c)** Two different output intensity curves for the cavity-only sample by pumping with 45 μJ and 330 μJ. **(d)** Output pumping curves for the CENA-2 and cavity-only devices.

As a receiver, the optical antenna array transfers efficiently free propagating light into the subwavelength cavity. According to the reciprocal theorem, the antenna is also a good reflector radiating localized emissions efficiently as free-propagating directional waves[33] (Fig.5a-b). Confocal scanning microscopic images (Fig.5c-d) indicate the radiation directivity of cavity-only and CENA. The divergence angle is 8° for CENA and ~43° for cavity-only (Fig.5d). The presence of the nanoantenna array has focused the radiation from the CENA device within a narrow angular span (Fig.5d). The nanoantenna array hence functions as a collector of pumping light, and also as a part of the plasmonic nanocavity; its configuration should therefore be carefully optimized. As an example, different nanoantenna heights of 5 nm, 15 nm, and 30 nm were set in calculating the local light intensity within the cavity; the result suggests that higher antennas transfer more pump-beam light into the cavity (Fig.5e). However, higher antennas also delocalize the cavity mode in the antennas (Supplementary Fig.S5), thereby increasing cavity loss. There is a similar trade-off between antenna and cavity modes in selecting antenna periodicity (Supplementary Fig.S6). We emphasize that low threshold can only be achieved when the nanocavity and the optical antennas are well matched with optimized device parameters including antenna periodicity (P), diameter of the fluorescence polystyrene sphere (D), and thickness of the metal film (T). For the experiments, we fabricated eight samples with different P, D, and T; lasing measurements indicate that the lowest threshold of 7 μJ is achieved at 65 nm, 350 nm, and 150 nm (Fig. 5f).

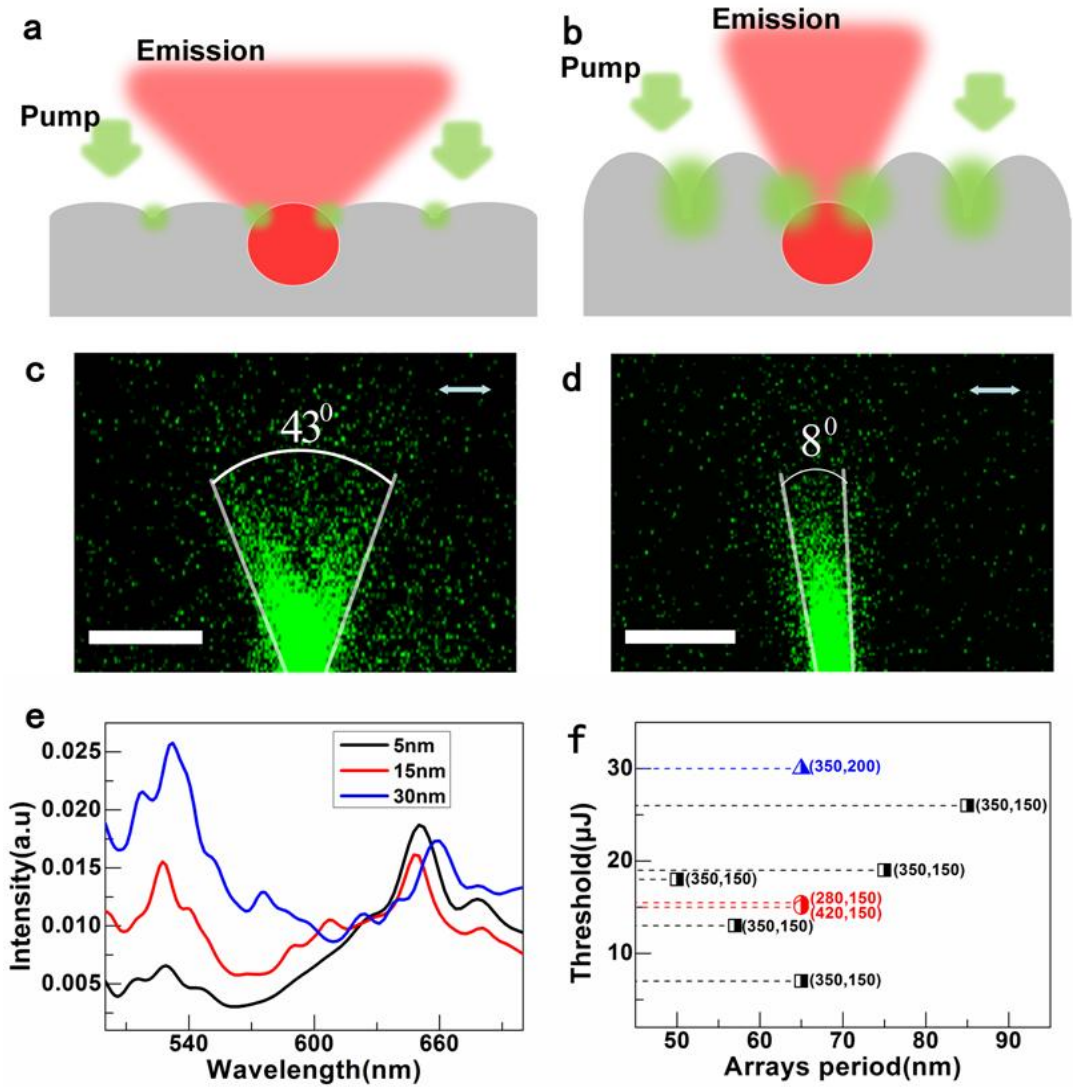

**Figure 5. (a-b)** Schematics of the trade-off mechanism between antennas and cavity. **(c-d)** XZ confocal scanning images of fluorescence sphere emissions from the CENA and cavity-only devices; scale bars located in the left-hand corner are 1 μm in length and the excitation laser polarization is along the *x*-axis (right corner of these two images). **(e)** Near-field intensity spectra for CENA with 3 different antenna heights (5 nm, 15 nm, and 30 nm). **(f)** Statistics for the thresholds in following CENAs with device parameters P, D and T.

The ultrahigh lasing threshold is still the main obstacle hindering the application of spaser[8-20], especially for the truly 3D-subwavelength-confined spaser[8, 9]. Combining the merits of an optical antenna array[33] and a plasmonic nanocavity[16, 17], a CENA device can operate as a low-lasing-threshold spaser. The performance of a CENA device can be improved if samples are fabricated with advanced nanomanufacturing techniques such as Focused ion beam (FIB) and e-beam to fashion the cavity and nanoantenna more accurately.

In summary, we have demonstrated plasmonic lasing in a nanocavity-embedded nanoantenna array fabricated by an effective and low-cost chemical method. The lasing threshold is lower than other room-temperature arrayed plasmonic nanolasers[18]. Combined with a nanoantenna array, we found that pumping efficiency is another important aspect in helping to lower the experimental lasing threshold of the spaser. The CENA structure provides a new platform in which to combine the merits of a nanocavity (high local density of states) and nanoantennas (effective transference of near-field and far-field optical energy) that will have potential in developing nano-scale light sources for photonic-integrated circuits and quantum information technology devices.

## AUTHOR INFORMATION

**Corresponding Author**

※ E-mail: yhlu@ustc.edu.cn or wangpei@ustc.edu.cn

**Author Contributions**

Y.L. and P.W. conceived the idea and supervised the project, C.Z., M.L., L.M., and C.L. perform the experiment, C.Z. and Y.N. do the simulation. C.Z. and Y.L. wrote this manuscript. All authors contributed to the discussions about this work, analysis of experimental data, comments on the manuscript and writing of the manuscript.

**Notes**

The authors declare no competing financial interest.

## ACKNOWLEDGEMENTS

This research is supported by National Key Basic Research Program of China (2012CB921900, 2012CB922003), Key Program of National Natural Science Foundation of China (61036005) and National Natural Science Foundation of China (61377053, 11274293, and 11374286). The authors thank Houqiang Jiang, Zheng Xi, Qiang Fu, Yong Wang, Erchan Yang, Yikai Chen, Kang Jiang and Xianghao Zeng for their helpful discussions.

# Supplementary Information

**EXPERIMENTAL METHODS**

**1. Sample fabrication**

The CENA device was fabricated using the following procedure: First, a mixed solution of perchloric acid with ethanol (volume ratio of 1:3) was prepared for polishing aluminium foil electro-chemically under a constant current of about 0.9 A at 0–4 ℃ for approximately 3 min. The surface roughness of aluminium foil was under 3 nm over an area of 3 μm$^2$. The hexagonally ordered anodic aluminium oxide (AAO) template was fabricated through a two-step anodizing process[1]. Second, fluorescence polystyrene particles (R-150 and R-200, Duke Scientific, Palo Alto, California, USA) were diluted by deionized water to prepare a monodispersed particle solution. These polystyrene particles were then dropped uniformly onto the surface of the AAO template and dried naturally. Third, a silver film was thermal-evaporated onto the surface of the AAO template, and then gently lifted off with double-side tape[2].

The cavity-only device was fabricated by performing the same procedure but replacing the AAO template with a clean glass substrate.

**2. Lasing measurements**

Lasing spectra were recorded using an automated imaging spectrometer (Jobin Yvon IHR 550, resolution for 0.025 nm) with integrating time of 1 s. A commercial Nd: YAG (Quantel, YG980, repetition rate ~10 Hz, pulse width ~10 ns) pulsed laser operating at 532 nm centre wavelength was employed to pump the sample at room temperature. An objective lens (40×, NA=0.6, Olympus) was used to focus the pump laser to a 20-μm diameter spot normal to the plane of the sample, the output emission light was recorded using the same objective lens. A red glass filter (R550) was used to remove scattered pump-beam light. A detailed schematic of the experimental setup is presented in Supplementary Fig.S8.

**3. Far-field reflection measurements**

Reflection spectra were recorded using a local spectral detection setup[3]. An oil immersion objective lens (100×, NA=1.45, Olympus) was used to focus white light onto our samples to a spot of diameter around 8 μm. All the reflection spectra were normalized to the spectra of quartz glass.

**4. Confocal scanning microscopy measurements**

A 488 nm Ar$^+$ ion laser was used to excite the fluorescence sphere embedded in either the nanoantenna array (CENA device) or slab silver film (cavity-only device). The XZ plane of the emission beam was captured using a confocal scanning microscopy imaging system (FV300 X71, Fluoview). A high numerical aperture objective lens (100×, NA=1.45, oil, Olympus) was used to ensure the resolution is high enough to establish the radiation directivity. The fluorescence emission light was collected after passing through a 565-nm filter.

**NUMERICAL METHODS**

**FDTD simulations**

We performed 3D FDTD simulations of our lasing system using commercial software (FDTD Solutions, Lumerical Solutions Ltd.). A theoretical model was configured to aid analysis and to explain the lasing mechanism, (Fig. 1a). In this case, the periodicity of the silver (Ag material, Johnson & Christy) antenna array was set to 65 nm (the diameter of antenna is 60 nm) and the antenna height to 15 nm; the fluorescent polystyrene spheres (diameter 160 nm, refractive index 1.59) were embedded in the antenna array; the spheres protruded 5 nm above the surface into air as dictated by the fabrication process). Simulations were performed under perfectly matched layer boundary conditions with plane-wave excitations incident normal to the *xy*-plane of the sample in the *z*-direction. The mesh size for the antenna-array region was set to 2 nm×2 nm×2 nm, and 4 nm×4 nm×4 nm for the nanocavity region.

**SUPPLEMENTARY FIGURES**

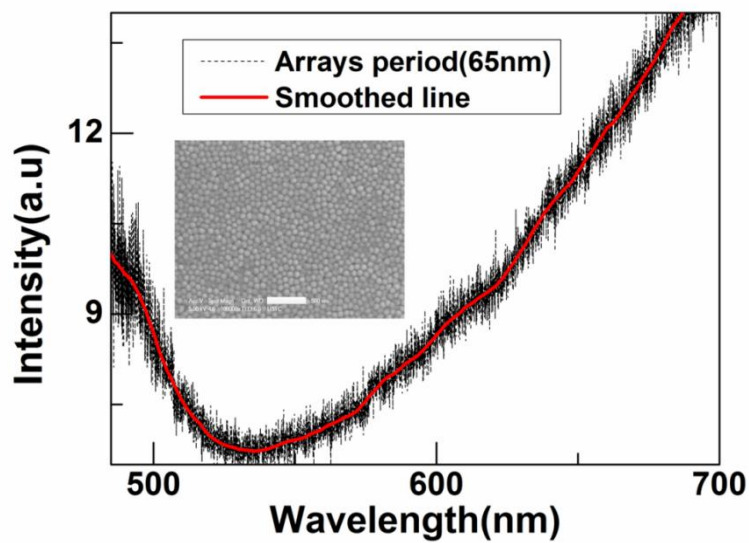

**Supplementary Figure S1. Reflection spectrum for the sample only contains antenna arrays with period of 65nm.** It illustrates that the absorption dip is around 530nm, which means the 532nm pumping laser can be converted to localized surface plasmon. The inset shows the SEM graph from top view, and the scale bar is 500nm.

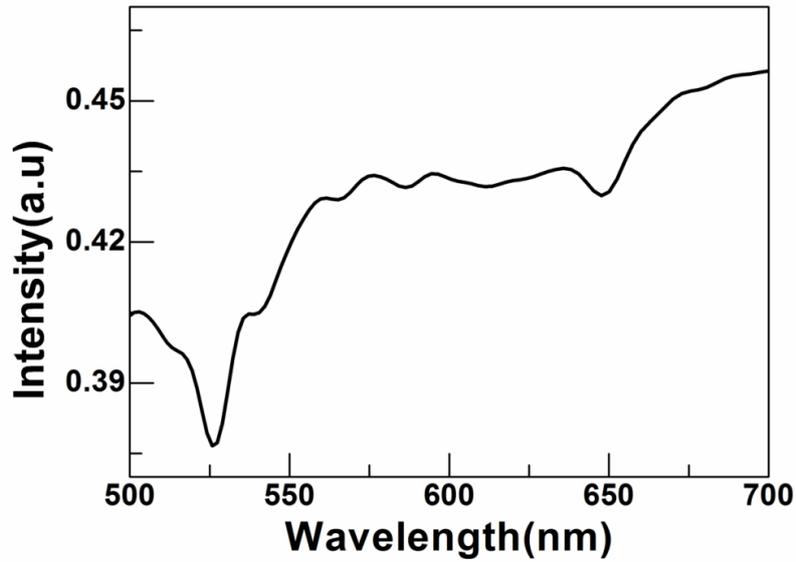

**Supplementary Figure S2. Calculated far field reflection spectrum of CENA.** The spectra were calculated by FDTD simulations with the simulation parameters as the Method mentioned. The spectra were collected in XY plane $2\mu m$-above the sample surface. According to this spectra, two resonance valleys located at 526nm (antenna mode) and 647nm (cavity mode) were observed, which agrees well with the experimental results.

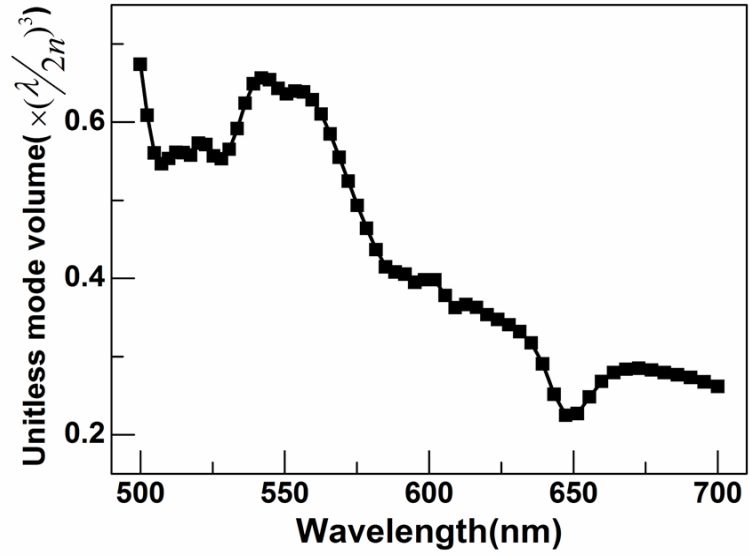

**Supplementary Figure S3. Mode volume for CENA.** The mode volume of CENA was calculated using a 3D finite difference time domain method (FDTD Lumerical solutions Inc.). Structure parameters were the same as that of Figure.S2. A 3D field profile monitor was set within a cubic of $200nm \times 200nm \times 200nm$ enclosing the entire nanocavity. Effective mode volume for this metallic resonator is defined as[4]

$$V = \frac{\int e(r)|E(r)|^2 d^3r}{\max\langle e(r)|E(r)|^2 \rangle}$$

where, $e(r)$ is the dielectric constant at position $r$, and $|E(r)|^2$ is the field intensity. Unitless mode volume $V_{mode}$ as a function of wavelength for CENA is achieved. The mode volume of the cavity mode at 647nm shrunk to $0.22(\lambda/2n)^3$, which was 4.3 fold smaller than that for the physical volume of the nano-cavity.

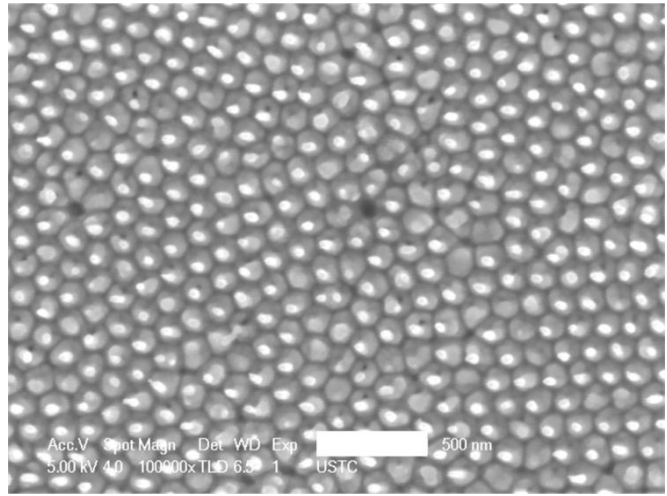

**Supplementary Figure S4. Top-view SEM of CENA-2.** The period for antenna arrays is around 155nm with height about 90nm, and the diameter of the facet of the nano-cavity is approximately 100nm, which is almost the same as that of the CENA sample present in Fig.1b.

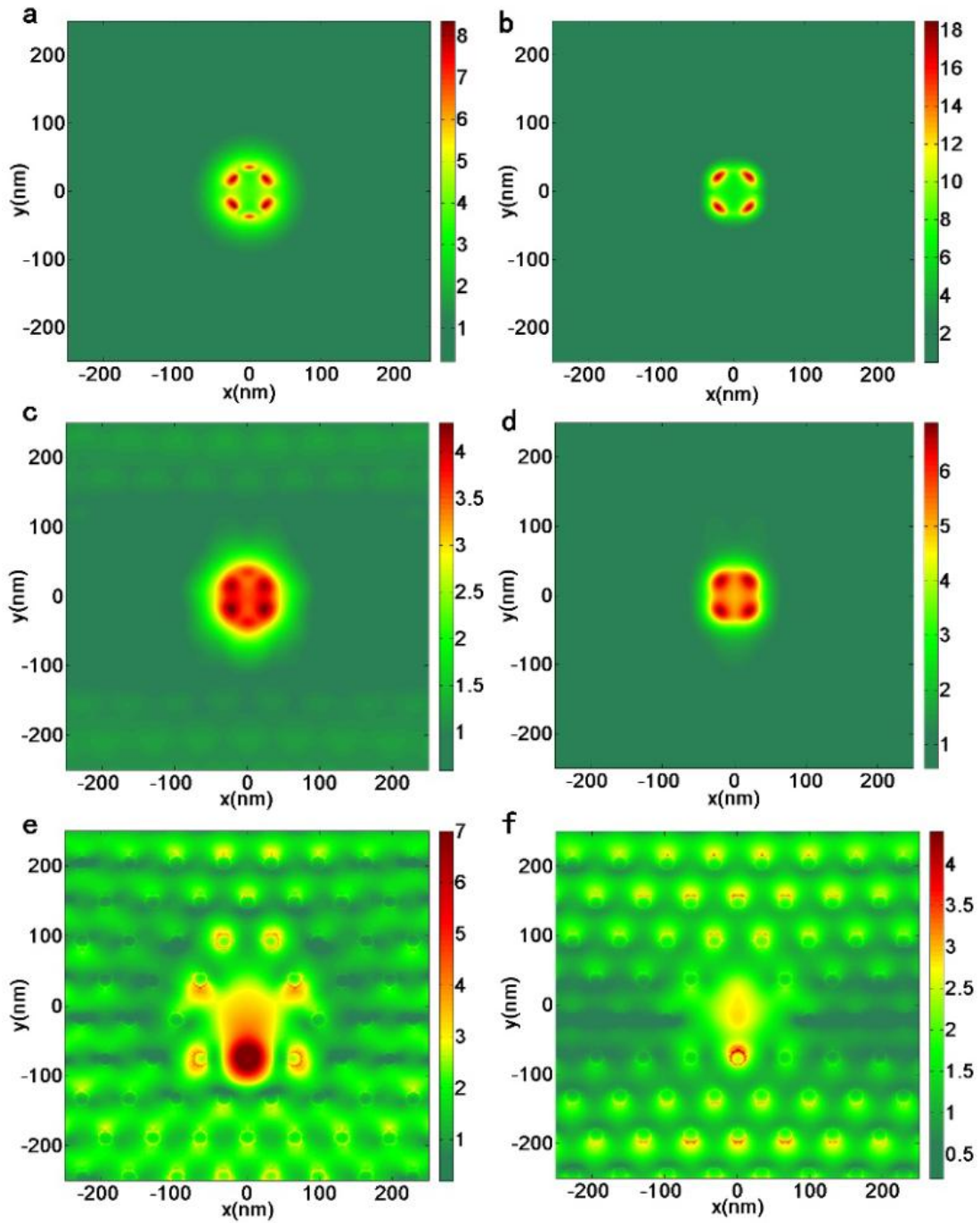

**Supplementary Figure S5. The effect of antenna: the Electromagnetic Field distribution of antenna and cavity mode. (a-b)** Electromagnetic field distributions of 532nm and 645nm(cavity mode) are presented for cavity-only device. **(c-d)** Electromagnetic field distributions of 530nm (antenna mode) and 645nm (cavity mode) are illustrated for CENA with antennas height of 5nm. As shown in Fig.S5c,

the coupling between antenna array and cavity is weak, almost the same as that of cavity-only device (Fig.S5a), so the pumping efficiency is very low. On the other hand, the cavity mode is localized very well in this case (Fig.S5d), which means less radiation loss. **(e-f)** Electromagnetic field distributions of 532nm (antenna mode) and 655nm (cavity mode) are given for CENA with antennas height of 30nm. In this case, the antenna mode transfer lots of pump energy into the nanocavity (Fig.S5e), but the cavity mode is delocalized by the higher antennas, thus lead to more radiation loss. Therefore, the height of antenna arrays has to be carefully optimized to achieve a low loss CENA plasmonic lasing due to the trade-off mechanism between the antenna mode and cavity mode.

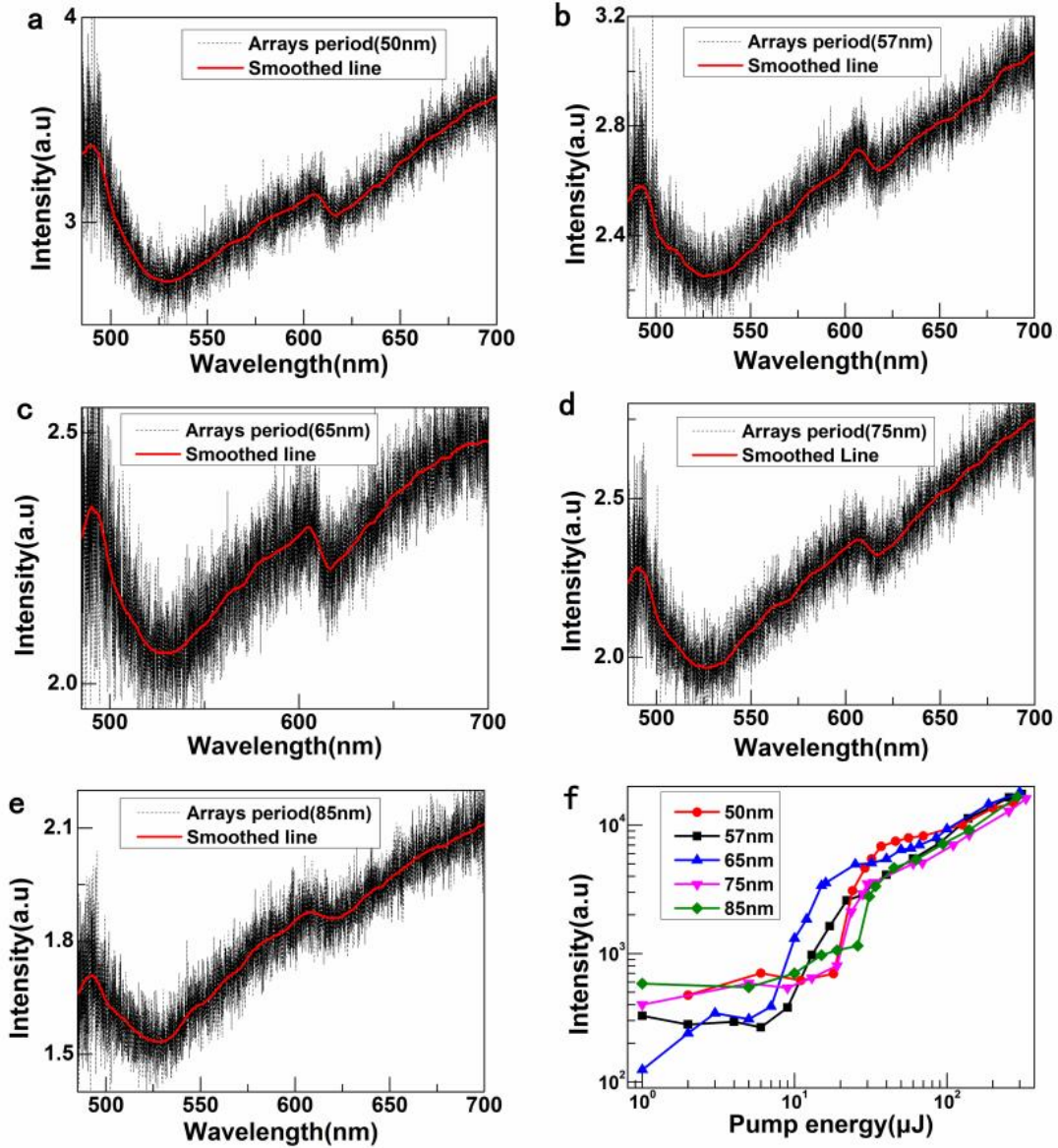

**Supplementary Figure S6. Reflection spectra of CENAs and the corresponding threshold curves (log-log) (a)-(e)** Far field reflection spectra for CENAs with antennas period from 50nm to 85nm (same diameter of 150nm of fluorescence sphere and silver film thickness about 350nm). According to Method of fabrication, the period and height of nanoantenna arrays are both governed by AAO template under different anodization voltages, and a small period could lead to a smaller height for antenna arrays. By increasing antennas period from 50nm to 85nm, their heights are growing from 7nm to 20nm (statistics average value). According to the trade-off

mechanism for antenna and cavity, it is noticeable that the cavity mode weakens as the height of nanoantennas grows. **(f)** Output-pumping curves are given for the five cases above. The plasmonic lasing threshold is the lowest when the antennas period is 65nm.

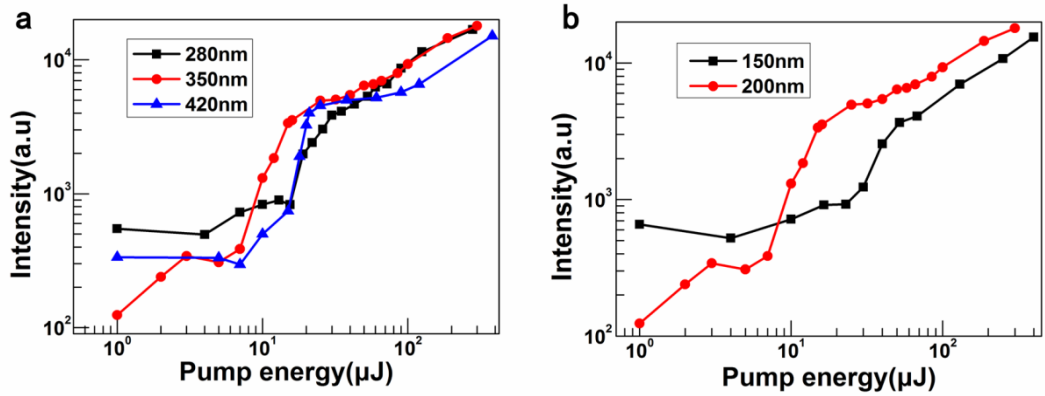

**Supplementary Figure S7.** (**a**) Output-pumping curves are given under three different film thickness of 280nm, 350nm and 420nm (same diameter of 150nm of fluorescence sphere and antennas' period of 65nm). With thickening silver films, the nano-cavity would be shaped a smaller opening solid angle exposed to air. It is easy to understand that the bigger open angle, the weaker pumping coupling between antennas and embedded cavity; while the smaller one, the higher radiation loss. (**b**) Output-pumping curves are given under two different diameter of fluorescence sphere (same period of 65nm of antenna arrays, same film thickness of 350nm).

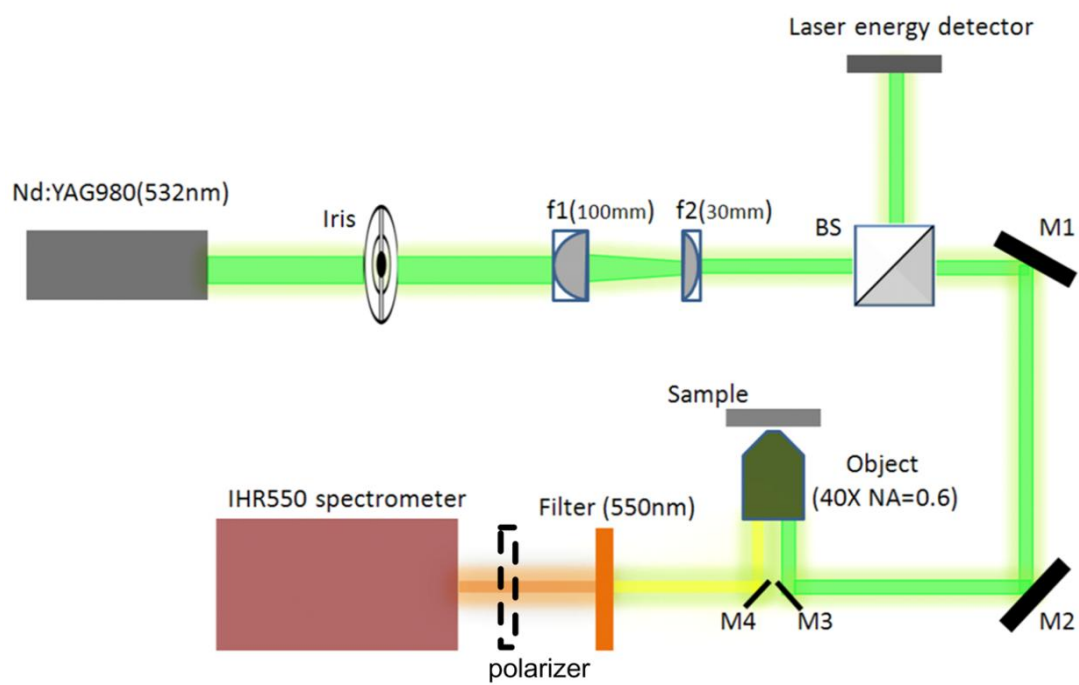

**Supplementary Figure S8| Lasing measurement setup**

A collimated pump light was focused down by a pair of lens (focal lengths of f1 and f2 are 100mm and 30mm respectively). A beam splitter (1:1) was used to divide the beam into two branches, one for the pump energy detection and the other for giving pump refuel to our sample. A polarizer was put in front of the spectrometer when we measure the lasing polarization state of CENA and cavity-only devices.